\documentstyle[aps,prb]{revtex}
\input epsf

\newcommand{\be}{\begin{equation}}
\newcommand{\ee}{\end{equation}}
\begin{document}

\title{
\begin{flushleft}
{\footnotesize BU-CCS-941003}\\
{\footnotesize {\tt comp-gas/9508001}}\\
{\footnotesize To appear in {\it Journal of Statistical Physics} (1995)}
\end{flushleft}
Simulating Three-Dimensional Hydrodynamics on a Cellular-Automata Machine
}
\author{
Christopher Adler\\
{\small Department of Earth, Atmospheric, and Planetary Sciences}\\
{\small Massachusetts Institute of Technology}\\
{\small Cambridge, MA \ 02139}\\
[0.3cm]
Bruce M. Boghosian \\
{\small Center for Computational Science}\\
{\small Boston University}\\
{\small 3 Cummington Street}\\
{\small Boston, MA 02215}\\
[0.3cm]
 Eirik G. Flekk\o y\footnote{
Current address:
Laboratoire de Physique M\'ecanique des Milieux Het\'erog\`enes,
Ecole Sup\'erieure de Physique et Chimie Industrielle,
10 rue Vauquelin,
75005 Paris France}\\
{\small Department of Earth, Atmospheric, and Planetary Sciences}\\
{\small Massachusetts Institute of Technology}\\
{\small Cambridge, MA \ 02139}\\
[0.3cm]
Norman Margolus \\
{\small Laboratory for Computer Science}\\
{\small Massachusetts Institute of Technology}\\
{\small Cambridge, MA 02139}\\
[0.3cm]
Daniel H. Rothman\\
{\small Department of Earth, Atmospheric, and Planetary Sciences}\\
{\small Massachusetts Institute of Technology}\\
{\small Cambridge, MA 02139}\\
[0.3cm]
}
\maketitle

\begin{abstract}
We demonstrate how three-dimensional fluid flow
simulations can be carried out on
the Cellular Automata Machine 8 (CAM-8),
a special-purpose computer for cellular-automata computations.
The principal algorithmic innovation is the use of a
lattice-gas model with a 16-bit collision operator that is
specially adapted to the machine architecture.
It is shown how the collision rules can be optimized
to obtain a low viscosity of the fluid.
Predictions of the viscosity based on a Boltzmann approximation
agree well with measurements of the viscosity made on CAM-8.
Several test simulations of flows in simple
geometries---channels, pipes, and a cubic array of spheres---are carried out.
Measurements of average flux in these geometries compare well with
theoretical predictions.
\end{abstract}
\vspace{0.3cm}


\section{Introduction}

A cellular automaton is a system of discrete variables on a lattice
which is updated according to some simple and local
rule~\cite{wolfram86a,toffoli87}.  Though applications of such models
range from physics to biology to social science~\cite{weisbuch91}, one
of the most exciting areas of interest in recent years has been
hydrodynamics.  This interest derives from the discovery by Frisch,
Hasslacher, and Pomeau in 1986 that discrete cellular-automaton models
of fluids, known specifically as {\em lattice-gas automata}, may be
constructed for the numerical solution of the Navier-Stokes
equations~\cite{frisch86,frisch87}.  Since their introduction, lattice
gases have been used to study a variety of problems in hydrodynamics.
Perhaps the greatest interest in the method is for the simulation of
problems that either involve complex boundaries, such as porous
media~\cite{rothman88a}, or complex fluids, such as
suspensions~\cite{ladd88,vanderhoef90} or immiscible
mixtures~\cite{rothman88b,appert90,rothman94}.

As in most other endeavors in computational physics, there is a great
need for fast, low-cost simulations.  Indeed, it has long been
recognized that the simplicity of lattice gases allows them to be
simulated on special-purpose hardware, called ``cellular-automata
machines''~\cite{margolus86,clouqueur87,toffoli87}.  Such machines
have been constructed for the simulation of general cellular automata,
and have been applied to a variety of systems.  They combine the
performance of supercomputers with the hardware simplicity of a
personal computer or workstation for these particular applications.

In this paper we describe an implementation of a three-dimensional
lattice-gas model~\cite{dhumieres86a} on a new cellular-automata
machine called CAM-8~\cite{margolus90,margolus95}.  CAM-8 retains the
high performance-to-cost ratio of previous cellular-automata machines,
but with considerably increased flexibility.  It is precisely this
flexibility which makes it attractive for simulating lattice gases.

Lattice gases (as well as other cellular automata) may be simulated by
the application of a lookup table, which gives an output state from the
input state at every site of the lattice.  The size of this table
increases with the number of degrees of freedom per site.  For this
reason lattice gases for three-dimensional
hydrodynamics~\cite{dhumieres86a}, which typically require 24 bits of
state (and hence $2^{24}$ possible states) per site, have proved
challenging to implement in situations where memory is limited, such as
on a distributed-memory multiprocessor.  Following the pioneering work
of H\'{e}non~\cite{henon87a,henon87b,henon89}, a considerable reduction
of table size was achieved by Somers and Rem~\cite{somers92}.  Precisely
what algorithm and collision table to use, however, remains a
machine-dependent question.  Because the CAM-8 allows one to work most
efficiently with units of 16 bits of state per site at any instant of
time, we have chosen to design a new collision strategy built upon
successive 16-bit table lookups.  Our results, in terms of minimization
of viscosity (and thus maximization of efficiency for the types of
simulations that interest us) are roughly comparable to those obtained
by previous workers.

The outline of the paper is as follows. We first briefly summarize the
principal architectural features of the CAM-8.  We then review the key
features of the face-centred hypercubic (FCHC) lattice-gas model, upon
which our implementation is based.  Next,
we show how the collision step of the FCHC lattice gas can be broken
down into operations that involve no more than 16 bits at a time.  We
also show how the collision rules can be optimized in a simple way to
give a small viscosity.  We compare our theoretical predictions of the
viscosity to results obtained by numerical simulation on CAM-8.  Finally,
we report a number of test simulations of flow through channels, pipes,
and a periodic array of spheres, and compare our results to analytic
predictions.  Our goal in this latter exercise is to provide some
practical guidelines for the application of our methods to the
measurement of the permeability of disordered media.

\section{CAM-8}

CAM-8 is an indefinitely scalable parallel computer optimized for the
large-scale simulation of three-dimensional cellular automata (CA)
systems.  It emphasizes simulation size, flexibility, and cost
efficiency, rather than ultimate performance---this emphasis leads to
a {\em virtual processor} design\cite{margolus95}.

\subsection{Hardware overview}

\begin{figure}[t]
\epsffile{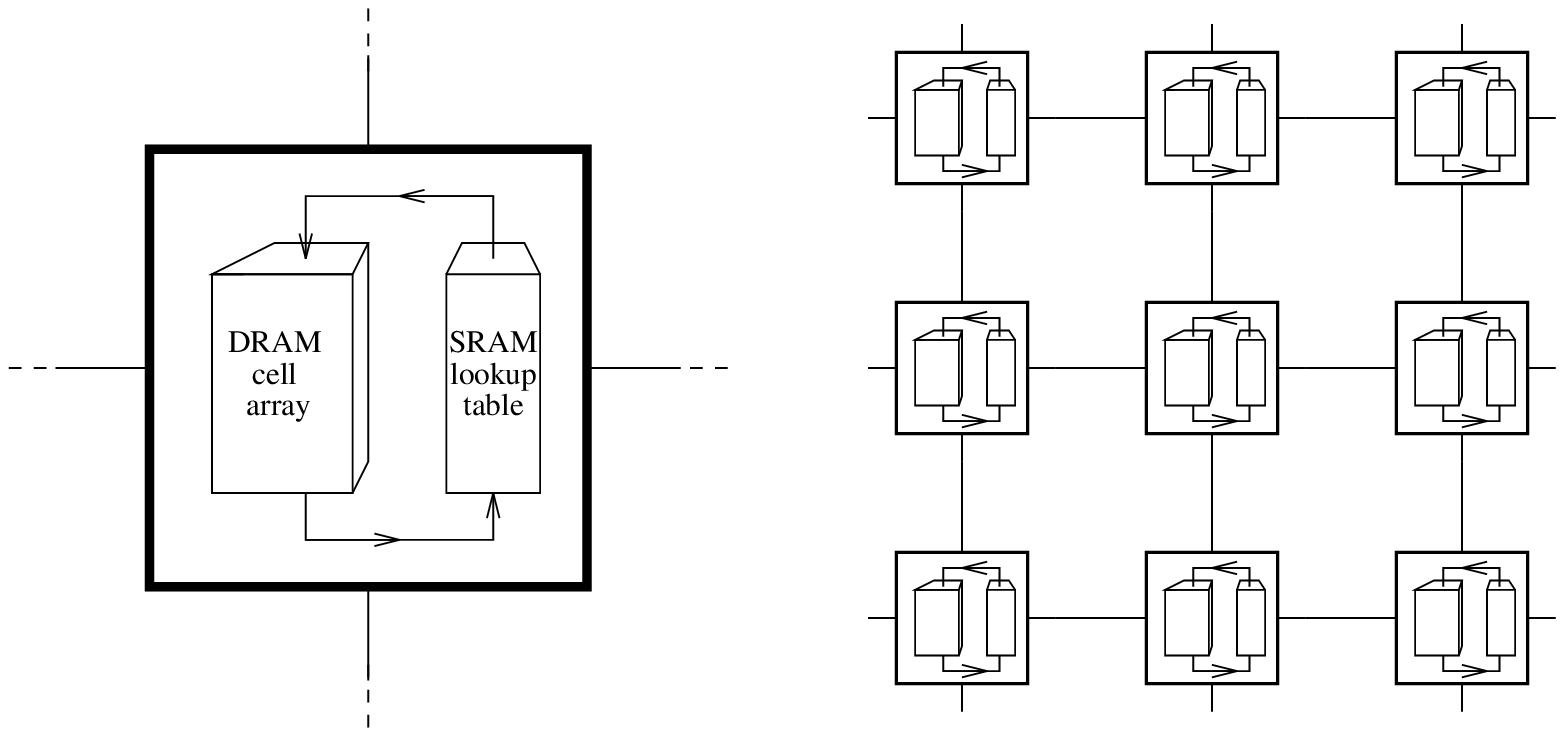}
\caption{ CAM-8 system diagram. On the left is a single processing node with
Dynamic Random Access Memory (DRAM) site data flowing through a Static
Random Access Memory (SRAM) lookup-table.
On the right is a spatial array of CAM8 nodes.}
\label{fig1}
\end{figure}

In CAM-8, each processor simulates the operation of up to millions of
spatial sites, with the site data stored in conventional DRAM-memory
chips.  This DRAM is scanned sequentially in an addressing pattern
that maximizes the speed of access, and data is read, updated (by
SRAM-memory table lookup) and put back.  This update cycle, which is
illustrated in Figure~\ref{fig1}a, resembles the operation of a video
framebuffer, and indeed CAM-8 is genetically more closely related to
framebuffer hardware than to conventional microprocessors.

Figure~\ref{fig1}b schematically illustrates an array of CAM-8
processing nodes (the array is actually three-dimensional in the
machine).  A uniform spatial calculation is handled in parallel by
these nodes, with each node containing an equal ``chunk'' of the
space.  In the diagram the solid lines between nodes indicate a local
mesh interconnection, used for spatial data movement between
processors.  Since each processor handles millions of spatial sites,
but only one at a time, these mesh wires are time-shared among
millions of sites.  The consequent reduction in interconnect, compared
to what would be needed in a fully parallel machine, makes large
three-dimensional simulations practical.  There is also a
communications network (not shown) connecting all nodes to a front-end
workstation that controls the CAM-8 machine.  The workstation uses
this network to broadcast simulation parameters to all or some of the
processing nodes.

Although the initial CAM-8 chip design accomodates machines with many
thousands of processing nodes, the first prototypes had only 8 nodes.
All of the simulations discussed in this paper were performed on such
an 8-node prototype, which has about the same amount and quality of
hardware as one might find in a low-end workstation: 2 Mbytes of SRAM,
64 Mbytes of DRAM, about two million gates of 1.2 micron CMOS logic,
with the whole thing running at 25 MHz.

\subsection{Programmable resources}

CAM-8 is a virtual-processor simulator of a fully parallel CA space.
Its virtual nature makes it possible to reconfigure and redirect its
computing resources, allowing the programmer to directly control
parameters such as

\begin{itemize}
\item
Number of dimensions
\item
Size and shape of the space
\item
Number of bits at a site
\item
Directions and distances of data movement
\item
Rules of data interaction (lookup tables)
\end{itemize}

The most novel facet of CAM-8's operation is the data movement, which
is lattice-gas-like.  Corresponding bits, one from each spatial site,
constitute a {\em bit-field}: each bit-field can be shifted in any
direction by a chosen amount.  The direction is chosen independently
for each bit-field, and the amount can be quite large---up to a few
thousand positions.\footnote{Such large spatial shifts are useful in
generating the high-quality random variables that are needed by
statistical mechanical simulations.}  Data movements are uniform
across the entire space---the hardware hides the fact that the space
is divided up among separate processor nodes.

The hardware only allows 16 separate bit-fields to be manipulated at a
time.  The 16 bits that ``collide'' at a given spatial site (i.e.,
land there due to their bit-field shifts) are replaced with a new
value given by a lookup table.  In fact, all data interaction is
performed by 16-input/16-output lookup tables.  Interactions that
involve more than 16 bits at each site must be synthesized as a
sequence of 16-bit interactions.

The lookup tables in the hardware are double-buffered: this means that
while one table is being actively used to scan all the sites and
update each in turn, the front-end workstation can broadcast the table
that will be used next.  Since the data movement is controlled by a
few pointers within each processor, these movements can be changed
quickly from one scan to the next.  Thus there is very little overhead
involved in applying a different lookup table {\em and} a different
set of bit-field movements at every scan of the space.

Finally, we note that most of our data analysis and data collection is
performed directly by dedicated statistics-gathering hardware.  The CA
space is split up into bins of a chosen size and lookup tables are
used to evaluate a function on each of the sites in each bin.  These
function data are collected by event counters and are continuously
reported back to the front-end as the simulation runs.

\section{The FCHC lattice gas}

Two fluids with quite different microscopic interactions may still have
the same macroscopic behavior. The reason for this is that the form of
the macroscopic equations of motion that describe this behavior depend
only on the conservation laws obeyed by these interactions and not on
their detailed form.  This was one of the main motivations for the
introduction of lattice-gas automata as a method to simulate fluid flow.

A lattice gas models a fluid as a large number of particles undergoing
simple interactions that conserve mass and momentum.  While the aim of
molecular dynamics is to simulate the real physics at the molecular, or
microscopic, scale, the microworld of lattice gases is fictitious.
Nevertheless, realistic macroscopic behavior is recovered when space and
time averages are performed.

One of the first lattice-gas models was introduced by Hardy, de Pazzis,
and Pomeau in 1976~\cite{hardy76}. This model, constructed on a square
lattice, succeeded in describing isotropically propagating sound waves,
and, if two particle species were introduced, diffusion.  The
hydrodynamics of this model, however, were anisotropic, as was the
damping of the sound waves.

The simplest and first lattice-gas model that produced isotropic
two-dimensional hydrodynamic behavior was introduced on a triangular
lattice by Frisch, Hasslacher, and Pomeau in
1986~\cite{frisch86,frisch87,doolen91,rothman94}.  The triangular
lattice is crucial for the isotropy of the flow dynamics, which,
technically, relies on the isotropy of the second and fourth order
tensors constructed from the basis vectors of the lattice.  In three
dimensions no regular lattice with this property exists.  However, the
four-dimensional face-centered hyper-cubic (FCHC) lattice has the
required symmetry~\cite{dhumieres86a}, as do its projections to three
dimensions.

\subsection{Description of the FCHC lattice}

The FCHC lattice is formed from the 4 basis vectors, $(\pm 1, \pm 1 , 0
,0)$, and the 20 additional vectors obtained by permuting the components
of these vectors. The FCHC lattice thus has a total of 24 basis vectors,
and we list these as follows:
{\footnotesize
\begin {equation}
\begin{array}{ccc}
{A}& {B}& {C} \\
\begin{array}{rrrrrr}
(&1& 0& 0& 1&) \\
(&1& 0& 0& -1& )\\
(& -1& 0& 0& 1&) \\
(& -1& 0& 0& -1&) \\
(& 0,& 1,& 1,& 0 &)\\
(& 0,& 1,& -1,& 0 &)\\
(& 0,& -1,& 1,& 0 &)\\
(&0,& -1,& -1,& 0 &)
\end{array}
&
\begin{array}{rrrrrr}
(& 0,& 1,& 0,& 1&) \\
(& 0,& 1,& 0,& -1& )\\
(& 0,& -1,& 0,& 1&) \\
(& 0,& -1,& 0,& -1&) \\
(& 1,& 0,& 1,& 0 &)\\
(& -1,& 0,& 1,& 0 &)\\
(& 1,& 0,& -1,& 0 &)\\
(& -1,& 0,& -1,& 0&)
\end{array}
&
\begin{array}{rrrrrr}
(& 0,& 0,& 1,& 1&) \\
(& 0,& 0,& 1,& -1&) \\
(&0,& 0,& -1,& 1&) \\
(& 0,& 0,& -1,& -1 &)\\
(& 1,& 1,& 0,& 0&) \\
(& 1,& -1,& 0,& 0 &)\\
(& -1,& 1,& 0,& 0& )\\
(& -1,& -1,& 0,& 0 &)
\end{array}
\end{array}.
\label{vectors}
\end{equation}
}
This grouping of the lattice vectors into three subgroups of eight will
be discussed further below.  For now we note that, if $\theta (n_A,m_B)$
is the angle between the $n$th vector of group $A$ and the $m$th vector
of group $B$, then $\theta (n_A,m_B) = \theta (n_B,m_C) = \theta
(n_C,m_A)$, and $\theta (n_A,m_A) = \theta (n_B,m_B) = \theta
(n_C,m_C)$, and each group contains 4 oppositely oriented pairs of
vectors.  In other words, the ordering of the velocity vectors is such
that group $B$ is related to group $A$ in exactly the same way as group
$C$ is related to $B$ and $A$ is related to $C$.

\begin{figure}[t]
\epsffile{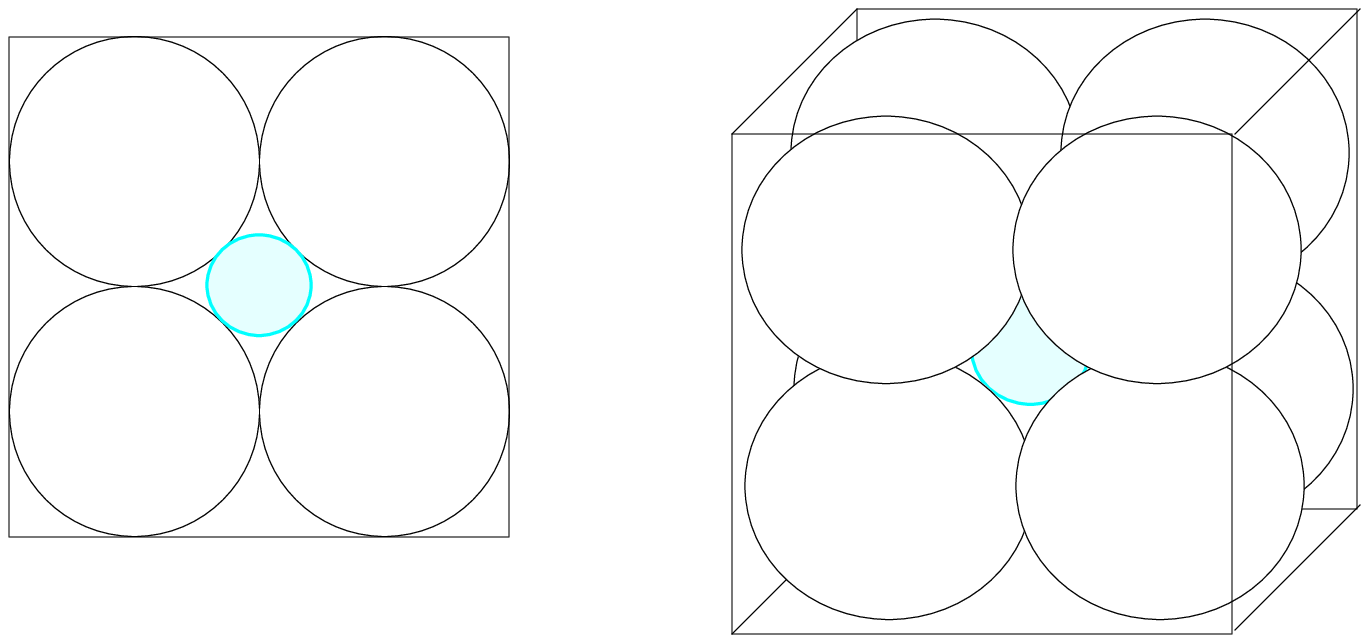}
\caption{{ Geometric method of visualization of the FCHC
lattice.} The ($2^2$, $2^3$) unshaded unit (circles, spheres) are packed
into a (square, cube) of side 4.  The shaded (circle, sphere) that is
tangent to them has radius equal to one less than the square root of the
number of dimensions.}
\label{fchcviz}
\end{figure}

It is possible to visualize the FCHC lattice by the geometrical
construction shown in Figure~\ref{fchcviz}.  Consider four unit circles
packed into a square of side 4 in two dimensions.  It is straightforward
to see that the inscribed circle, tangent to each of these four unit
circles, has radius $\sqrt{2}-1$.  In three dimensions, when eight unit
spheres are packed into a cube of side 4, the inscribed sphere has
radius $\sqrt{3}-1$.  Likewise, in $D$ dimensions, when $2^D$ unit
hyperspheres are packed into a hypercube of side 4, the inscribed
hypersphere has radius $\sqrt{D}-1$.  In particular, in four dimensions
($D=4$), the inscribed hypersphere is also a unit hypersphere.  This
nontrivial packing of unit hyperspheres provides an
alternative description, within a factor of $\sqrt{2}$,
of the FCHC lattice: the centers of the
hyperspheres are the lattice vertices, and tangent hyperspheres
correspond to linked vertices.

This geometrical description makes it clear that sixteen of a given
lattice site's neighbors lie on the corners of a perfect
four-dimensional hypercube, while the other eight neighbors correspond
to the eight faces of that hypercube\footnote{Recall that a
$D$-dimensional hypercube has $2D$ faces.}.  Presumably, this is how the
face-centered hypercubic lattice derives its name.  Note also that there
are three distinct ways in which this partition of the lattice vectors
into a group of sixteen and a group of eight may be carried out.  These
three ways correspond to letting each of the three subgroups in the list
(\ref{vectors}) be the group of eight.

To see this explicitly, note that the proper orthogonal matrix
\begin{equation}
\Lambda = \frac{1}{\sqrt{2}} \left(
\begin{array}{cccc}
+1&+1&0&0 \\ +1&-1&0&0 \\ 0&0&+1&+1 \\ 0&0&+1&-1
\end{array} \right),
\label{omat}
\end{equation}
when applied to each of the twenty-four lattice vectors in
equation~(\ref{vectors}), yields~\footnote{For brevity, we have omitted the
normalization factor $1/\sqrt{2}$ in front of each of these vectors.}
{\footnotesize
\begin{equation}
\begin{array}{ccc}
{A}& {B}& {C} \\
\begin{array}{rrrrrr}
(&+1,&+1,&+1,&-1&)\\
(&+1,&+1,&-1,&+1&)\\
(&-1,&-1,&+1,&-1&)\\
(&-1,&-1,&-1,&+1&)\\
(&+1,&-1,&+1,&+1&)\\
(&+1,&-1,&-1,&-1&)\\
(&-1,&+1,&+1,&+1&)\\
(&-1,&+1,&-1,&-1&)
\end{array}
&
\begin{array}{rrrrrr}
(&+1,&-1,&+1,&-1&)\\
(&+1,&-1,&-1,&+1&)\\
(&-1,&+1,&+1,&-1&)\\
(&-1,&+1,&-1,&+1&)\\
(&+1,&+1,&+1,&+1&)\\
(&-1,&-1,&+1,&+1&)\\
(&+1,&+1,&-1,&-1&)\\
(&-1,&-1,&-1,&-1&)
\end{array}
&
\begin{array}{rrrrrr}
(& 0,& 0,& +2,& 0&)\\
(& 0,& 0,& 0,& +2&)\\
(& 0,& 0,& 0,&-2&)\\
(& 0,& 0,&-2,& 0&)\\
(& +2,& 0,& 0,& 0&)\\
(& 0,& +2,& 0,& 0&)\\
(& 0,&-2,& 0,& 0&)\\
(&-2,& 0,& 0,& 0&)
\end{array}
\end{array}.
\label{vectors2}
\end{equation}
}
In this representation, it is manifest that the lattice vectors of
subgroups $A$ and $B$ lie at the corners of a perfect four-dimensional
hypercube, and that those of subgroup $C$ lie on its faces.  In what
follows, we therefore refer to this representation as the {\it explicit
hypercubic frame}.  This frame will be useful for describing the
isometries of the FCHC lattice in Section~\ref{ssiso}.

Finally, we note that isotropic three-dimensional hydrodynamics can be
obtained from three-dimensional projections of this four-dimensional
lattice.  Due to the staggered nature of the FCHC lattice, this requires
only two lattice spacings in the fourth dimension.  Alternatively (and
equivalently), we can simply project out the fourth coordinate of the
lattice vectors in equation~(\ref{vectors}).  The resulting set of
three-vectors are the lattice vectors of an irregular lattice in three
dimensions.  The neighbors of a point $(0,0,0)$ on this lattice can be
described in Cartesian coordinates as the six on-axis neighbors at
distance one ($(\pm 1, 0, 0)$, $(0, \pm 1, 0)$, and $(0, 0, \pm 1)$),
each counted with multiplicity two (double bonds), and the twelve
neighbors at distance $\sqrt{2}$ ($(\pm 1, \pm 1, 0)$ and permutations).
The four-tensor constructed by these lattice vectors is perfectly
isotropic.

\subsection{Microdynamics}

The two basic steps of the LGA are 1) propagation of the particles, and 2)
collisions. The particles reside on the lattice sites only, and there
can be at most one particle per direction at any given site and time.
The hydrodynamic behavior of the model depends on the fact that the
collision step conserves mass and momentum (conservation of these
quantitites holds trivially in the propagation step).

In order to introduce flow with solid walls present one must introduce
new collisions which prevent particles from moving across the
boundaries. These might be either of the bounce-back type which send
particles back into the direction from which they came, or of the mirror
reflection type where only one component of the particle's momentum is
changed.  When particle velocities are averaged the effect of the
bounce-back collision is a hydrodynamic no-slip boundary
condition~\cite{cornubert91,ginzbourg94}.

In order to introduce a body-force, like gravity, an additional
collision step is needed that puts momentum into the system.  This can
be done in several ways, one of which is to flip particle velocities at
a few randomly chosen sites into the direction of the forcing.

The state at a single site (at position $\mbox{{\bf x}}$ at time $t$) is
given by the 24 occupation numbers $n_i (\mbox{{\bf x}} , t)\in\{
0,1\}$, which are simply the particle number in direction $i$.
The time development of the $n_i$'s is given by the microdynamical
equation
\be
n_i(\mbox{{\bf x}} + \mbox{{\bf c}}_i , t+1) = n_i(\mbox{{\bf x}}, t)
+\Omega_i (\{ {\bf n}(\mbox{{\bf x}} ,t)\} ).
\label{Boltzmann}
\ee
The term $\Omega_i (\{ {\bf n}(\mbox{{\bf x}} ,t)\} )$, where ${\bf n} =
(n_1, n_2, \ldots , n_{24})$, is the change in $n_i$ due to collisions,
$\mbox{\bf c}_i$ are the velocity vectors connecting neighboring lattice
sites, $t$ is the time, and the time increment corresponding to a
combined propagation and collision step is unity.

The full detailed state of the lattice gas is given by the set of all
$n_i$'s on all sites. In simulations on a computer, all this information
is packed into 24 bits of information at each site, and is stored and
updated.  However, the quantities of physical interest are the (space
and/or time and/or ensemble) averaged mass and momentum densities.  On
the CAM-8 this averaging can be performed by the application of look-up
tables and read to the front end without loss of speed.

\subsection{The 16-bit collision operator and random isometries}
\label{ssiso}

\begin{figure}[t]
\epsffile{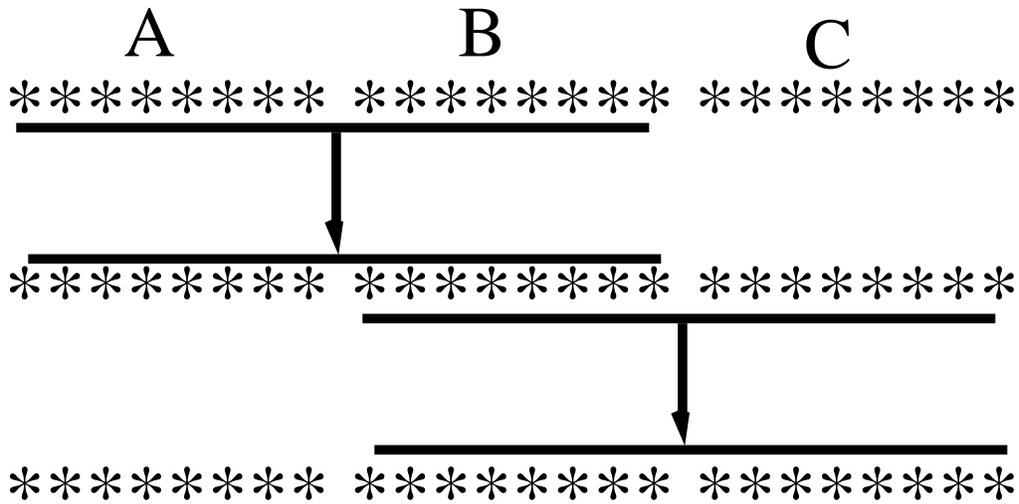}
\caption{{ The structure of the two step collisions.}  Each (*)
represents a particle occupation number (a bit). The arrows show which
bits are influenced by each collision step. \label{fig2}}
\end{figure}

In a three-dimensional lattice gas the number of possible collision
rules, represented by the collision operator $ \Omega$, is very large.
If one 24-bit output configuration for each of the $2^{24} = 16$M input
configurations were stored in a table, the table would have a size of at
least 48 Mbytes.  This is bigger than what the local SRAM memory in each
node of the CAM-8 permits, and it is also bigger than the local memory of some
massively parallel machines, like the Connection Machine CM2.  The
restriction for the CAM-8 is the 16-bit limit of the lookup tables.  In
order to adapt to this restriction each 24-bit collision is split into
two 16-bit collisions. The structure of this factorization is shown in
Figure~\ref{fig2}.  The particle velocities are split into the three
groups of eight introduced in equation~(\ref{vectors}).  The collisions,
which are given by a single table, involve 16 particles at a time and
act first on the first 16 particles (subgroups $A$ and $B$) and then on
the last 16 (subgroups $B$ and $C$). The overall mass and momentum of
the 16 particles is conserved at each step.

Such a division of the collision rule
into three fixed sets of spatial directions can introduce
anisotropies.  In order to prevent this effect, the particles are first
subjected to a random precursor transformation---an isometry $p$---before
the collision $C$; after the collision they are then subjected to
the inverse isometry $p^{-1}$.  The isometries $p$ form a group $G$
defined as the set of transformations that map the set of lattice
vectors $\{\mbox{\boldmath c}_i \}$ to itself, preserving the vectors'
lengths and the angles between them.  The total collision step can thus
be represented by the operator $p^{-1} C p$, where $C$ is deterministic
and is given by the 16-bit table, and $p$ is randomly chosen from the
isometry group $G$ at every timestep.  Note that $p$ can be the same for
all sites on the lattice.

It can be shown that there are altogether 1152 elements in
$G$~\cite{henon87b}, and each element is a map acting on all the 24
bits.  It is therefore crucial that the isometries can be factorized
into factors that, like $C$, decompose into operations on subsets of the
24 bits.

This factorization takes a convenient form when specified by the action
of the isometries on the coordinates of the velocity vectors in the
explicit hypercubic frame defined by equation (\ref{vectors}).
In this frame the
isometries take the form of a product of simple coordinate inversions and
permutations, known as $R$ and $P$ isometries, respectively,
combined with the forward or reverse cyclic interchange of
the subgroups $A$, $B$ and $C$~\cite{henon87a}.  Specifically, the
isometries corresponding to the forward ($(A,B,C)\rightarrow (C,A,B)$)
and reverse ($(A,B,C)\rightarrow (B,C,A)$) cyclic interchange of the
three subgroups are denoted by $S_1$ and $S_2$, respectively.  No tables
are needed for these operations.  The
$R$ and $P$ isometries are also particularly simple because they do
not map vectors between $A + B$ and $C$. Thus they can be encoded as
16-bit and 8-bit tables acting on $A + B$ and $C$, respectively.

As a concrete example, consider the axis-inversion isometry $R_4$,
implemented by negating the fourth coordinate in the explicit hypercubic
frame.  As before, we enumerate the lattice vectors in this frame by
$1_A,\ldots, 8_A, 1_B,\ldots, 8_B, 1_C,\ldots, 8_C$, where, for example,
$n_A$ denotes the $n$th entry of column $A$ in equation~(\ref{vectors2}).  It
is evident that the application of $R_4$ amounts to swapping the lattice
vectors $1_A$ and $5_B$, $2_A$ and $7_B$, $3_A$ and $6_B$, $4_A$ and
$8_B$, $5_A$ and $1_B$, $6_A$ and $2_B$, $7_A$ and $3_B$, $8_A$ and
$4_B$, and $2_C$ and $3_C$.  The orthogonal matrix, equation~(\ref{omat}),
then maps these swaps back to the corresponding swaps in the original
frame, equation~(\ref{vectors}), even though it is evident that $R_4$ is no
longer a simple axis inversion in that frame.  So, for example, ordering
the lattice vectors and their corresponding bits according to equation
(\ref{vectors}), we see that the state
\begin{equation}
101010101010101010101010
\end{equation}
maps to the state
\begin{equation}
110010101010110011001010
\end{equation}
under the isometry $R_4$.

To apply a general isometry $p$ to the lattice,
we exploit the fact that $p$ can be written
in the product form
\begin{equation}
\left( \begin{array}{c} I \\ S_1 \\ S_2 \end{array} \right)
\left( \begin{array}{c}I\\ R_1 \end{array} \right)
\left( \begin{array}{c}I \\ R_2\end{array} \right)
\left( \begin{array}{c} I \\ R_3\end{array} \right)
\left( \begin{array}{c}I \\ R_4 \end{array} \right)
\left( \begin{array}{c}I\\P_{12}\end{array} \right)
\left( \begin{array}{c}I\\P_{13}\\P_{23} \end{array} \right)
\left( \begin{array}{c} I\\P_{14}\\P_{24} \\P_{34}\end{array} \right),
\end{equation}
where the indices of the $R$ and $P$ isometries refer to the axis to be
inverted or the axes to be permuted, respectively. The elementary $R$
and $P$ isometries are their own inverses whereas $S_1$ is the inverse
of $S_2$.
To choose a random isometry at each time step,
eight random numbers are
computed and used to choose one isometry from each column above.
The resulting combined isometry is then performed by the successive
application of tables corresponding to each of the elementary
isometries. The inverse isometry is obtained by the application of the
tables in the reverse order.

\subsection{Memory usage and implementation of the algorithm}

The memory in the CAM-8 is arranged so that there are four 16-bit
subcells at each lattice site.  The initial organization of the memory
is to occupy the first eight bits of the first three subcells with the
three groups, $A$, $B$ and $C$, of lattice vectors, respectively. The
subcells are given the same labels, $A$, $B$ and $C$, as the velocity
groups they contain.  The extra bits are used as logical flags and swap
space which can minimize the number of steps required to implement the
collision algorithm, trading abundant memory for computational speed.
In addition, there is a subcell $D$ which contains flags that indicate
whether or not a lattice site is solid, and whether or not forcing is to
be performed (if possible) at that site.  Thus bits 0 through 7 of the
subcell $A$ indicate the presence or absence of particles moving in
lattice directions $A$ , while bits 0-7 of subcell $B$ indicate
similarly for lattice directions $B$, and bits 0-7 of subcell $C$
indicate similarly for lattice directions $C$.

The collision algorithm proceeds in five steps:
\begin{enumerate}
\item
Apply forcing and reorganize the memory to prepare subsequent steps.
\item
Apply randomly chosen $R$, $P$, and $S$ isometries.
\item
Apply collision table.
\item
Apply inverses of isometries applied in step 2 in reverse order.
\item
Reflect lattice vectors at solid sites and restore initial organization
of memory.
\end{enumerate}

Forcing is applied by adding momentum to a particle if possible and if
the flag at the site permits.  The bit plane of flags (or planes, if
forcing in more than one direction is done simultaneously) is randomly
shifted at each time step so that forcing is on average uniform over the
space.  Because the same lookup tables are necessarily applied at every
site, the amount of forcing is set in the initial conditions by setting
the number of positively set flags.  The amount of forcing applied
cannot be precisely tuned, since not all particle configurations can be
changed to increase the momentum in a given direction.  For this reason the
applied force must be measured along with the other observables.

At the solid sites all momenta are reversed. This is an operation that
can be performed on each of the three groups of bits separately, since
each group contains four pairs of oppositely oriented momenta.  The
momentum reversals are handled in the following manner.  The flag that
indicates whether or not a site is solid is copied into each of the
first three subcells.  If the site is solid, the momentum information in
each subcell (in bits 0-7) is moved to the higher eight bits (8-15) and
the lower bits are set to zero.  The subsequent steps in the collision
algorithm are thus applied to the eight bits of zero instead of the
momentum information, and the net result is no change.  At the end of
the collision step, the presence of zeros in the first eight bits of
each subcell indicate that the site is solid, and that the momentum
information in the high eight bits should be reflected through the
origin and repositioned in the lower eight bits of each subcell.  This
algorithm is robust in the case where the site is not solid but no
particles are present, since, in this case, the reflection of the bits
will cause no change.

When forcing is being applied in only a single direction, the collision
algorithm requires 26 applications of lookup tables when an $S$ isometry
is applied, and 14 applications when the $S$ isometry chosen is the
identity, which happens 33\% of the time.  Thus the average number of
lookup table applications required per time step is 22.  The performance
of the present algorithm on the 8 node machine (1 node handles 4M
16-bit sites ) is 7M site updates per second.  It benefits greatly from
the fact that the CAM-8 can download one table while being busy using
another, without loss of speed.

\section{Optimization of the viscosity}

Typically, lattice-gas collision rules are chosen to maximize
a ``Reynolds coefficient'' and thus the Reynolds number of a simulation.
For the application we envision---namely, slow flow through disordered
porous media---accuracy and efficiency of the simulation improve
as the kinematic viscosity is reduced \cite{ferreol94}.
In this section we show how we choose our 16-bit collision tables to
obtain a low viscosity.

By employing the Boltzmann approximation, which assumes that the
particles entering a collision are
uncorrelated, it is possible to derive a closed expression for the
kinematic viscosity $\nu$ in terms of the collision rules and the
average density $d \leq 1$ per direction $i$.
The formula, due to H\'{e}non~\cite{henon87b}, has
the form
\be
\nu =   \frac{1+Q}{6(1-Q)} \; ,
\label{viscosity}
\ee
where
\be
Q = \frac{1}{18} \sum_{ss'} \; A(s \rightarrow s') \; d^{m-1}
(1-d)^{23 - m} \; Y_{\alpha \beta} (s) Y_{\alpha \beta} (s')
\label{Q}
\ee
and $A(s \rightarrow s')$ is the probability that the state $s$
goes into $s'$.
The sum runs over all possible input and output states $s$ and $s'$,
where $s = \{ s_i \}$, $m = \sum_i s_i$ is the mass at a site,
${\alpha }$ and ${ \beta}$ are cartesian indices with summation
over repeated indices implied, and
\be
Y_{\alpha \beta} = \sum_{i=1}^{24} s_i \left(
c_{i\alpha}c_{i\beta} - \frac{1}{2} \delta_{\alpha \beta} \right)
\label{Y} \ee
where $\delta_{\alpha \beta}$ is the Kronecker delta.
The term $Y_{\alpha \beta} (s) Y_{\alpha \beta} (s')$
is easily evaluated as
\begin{eqnarray}
Y_{\alpha \beta} (s) Y_{\alpha \beta} (s') &=&
\sum_{ij} s_i s'_j   \left(
c_{i\alpha}c_{i\beta} -
\frac{1}{2} \delta_{\alpha \beta} \right)
 \left(
c_{j\alpha}c_{j\beta} -
\frac{1}{2} \delta_{\alpha \beta} \right) \nonumber \\
&=& \sum_{ij} s_i s'_j \left( ( \mbox{\bf c}_i
\cdot  \mbox{\bf c}_j )^2 -1  \right)  \nonumber \\
&=& \sum_{ij} s_i s'_j {\cal A}_{ij}
\end{eqnarray}
where
\be
{\cal A}_{ij}=( \mbox{\bf c}_i \cdot \mbox{\bf c}_j )^2 -1.
\ee
The angle between lattice vectors $i$ and $j$ different subgroups is
either $\pi /3$ or $2 \pi /3$, and $|{\bf c}_i|=\sqrt{2}$, so $\mbox{\bf
c}_i\cdot \mbox{\bf c}_j = \pm 1$, so ${\cal A}_{ij}=0$ for $i$ and $j$
in different subgroups.  It follows that the $24\times 24$ matrix ${\cal
A}$ is block diagonal,
\be
{\cal A} =
\left(
\begin{array}{ccc} A & 0 & 0 \\ 0 & A & 0 \\
0 & 0 & A \end{array} \right),
\ee
where the $0$'s denote $8\times 8$ null matrices, and the $8\times 8$
matrix of components $A$ for lattice vectors in the same subgroup is
given by
\be
A = \left(\begin{array}{cccccccc}
 +3 & -1 & -1 & +3 & -1 & -1 & -1 & -1 \\
 -1 & +3 & +3 & -1 & -1 & -1 & -1 & -1 \\
 -1 & +3 & +3 & -1 & -1 & -1 & -1 & -1 \\
 +3 & -1 & -1 & +3 & -1 & -1 & -1 & -1 \\
 -1 & -1 & -1 & -1 & +3 & -1 & -1 & +3 \\
 -1 & -1 & -1 & -1 & -1 & +3 & +3 & -1 \\
 -1 & -1 & -1 & -1 & -1 & +3 & +3 & -1 \\
 -1 & -1 & -1 & -1 & +3 & -1 & -1 & +3
\end{array} \right) \; .
\ee
Thus, heuristically speaking, collision events that take a particle from
one lattice vector to another---neither the original one nor its
negation---in the same subgroup of eight are the most preferred in
terms of minimizing $Q$ (score $=-1$).  Collisions that take a particle
to a different subgroup are next (score $=0$).  Finally, collisions that
take particles to themselves or their negation are least preferred
(score $=+3$).  The collision set that minimizes this scoring
is optimal.

Due to the symmetric ordering of the velocity vectors, $Q$ splits into a
sum of three terms
\be
Q = Q_A + Q_B + Q_C \; ,
\ee
corresponding to the group A, B and C respectively.  We consider only
the first two terms, which corresponds to a 16-bit table, and choose the
collisions that minimize $(Q_A + Q_B)$.  Note that $\nu = \nu (Q)$ is an
increasing function of $Q$, so that minimizing $Q$ is equivalent to
minimizing $\nu$.  Minimizing $(Q_A + Q_B)$, however, is not equivalent
to minimizing $Q_A + Q_B + Q_C$. The former quantity results from a
single application of the 16-bit table, whereas the latter quantity
results from applying the 16-bit table twice.  Only $Q_A$ remains the
same after the second application.

The reason for doing only the restricted optimization is twofold.
First, by restricting the optimization to the 16-bit space with $2^{16}$
velocities, an exhaustive search through all possible output states
corresponding to every input state is feasible and requires only
about one hour of computation time on a Sun Sparcstation 2.
By comparison,
an exhaustive search through all output states corresponding to the
$2^{24}$ possible input states of the complete collision table would
require increase computation time by a factor of $(2^8)^2 \approx 65000$.

Second, we argue that the process of minimizing $Q_A + Q_B$ approximates
the process of minimizing $Q$.  From the above equations it is seen that
minimizing $Q$ means finding output states that select the matrix
elements $-1$ in $A$. This corresponds to maximizing the number of
perpendicular directions between the velocities of the input and output
states within each subgroup of eight.  Physically it corresponds to
minimizing the momentum flux and thereby the viscosity.  For this
purpose it is advantageous to let a particle scatter into one of the
directions of its own subgroup which give a contribution $-1$ to the
matrix product, rather than into another subgroup, which would give no
contribution to the matrix product.  The 16-bit collisions certainly
allow for the particles to go into their own subgroups.

However, the selection of output states is restricted, not only by mass
and momentum conservation, but by the limited number of velocities in
the 16-bit states as well.  Also, since the 16-bit table is applied
twice to give the full collision, the eight particles that are involved
in both collisions may scatter back into a disadvantageous direction.
The probability that such backscattering occurs, however,
appears to be small enough to give a
relatively low viscosity.  However, a refined scheme that takes into
account that the 16-bit collisions are indeed applied twice would be expected
to decrease the viscosity further.

\begin{figure}[t]
\epsffile{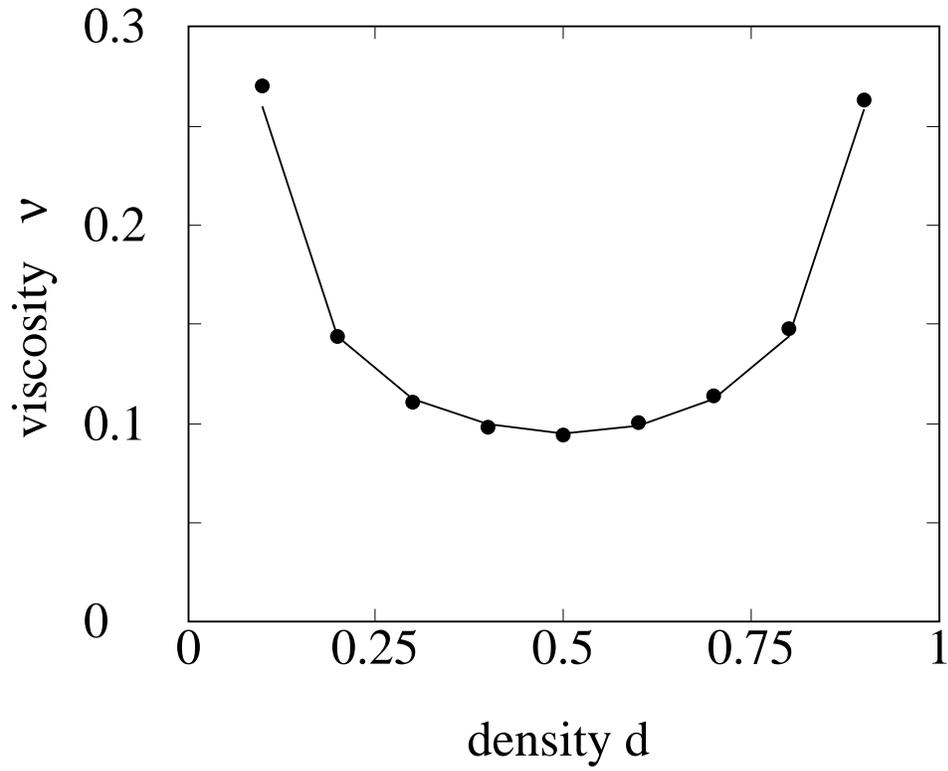}
\caption{{ Viscosity $\nu$ as a function of the reduced density $d$.}
The line shows the Boltzmann values and the dots show the values
obtained from simulations.\label{fig3}}
\end{figure}

Our optimized viscosity, determined from
the Boltzmann approximation of equations~(\ref{viscosity}-\ref{Y}),
is given as a function of the reduced
density $d$ in Figure~\ref{fig3}. Here it is also compared to measurements
of the viscosity by simulation of a relaxing shear wave on the CAM-8.
Specifically, simulations were initialized with the velocity profile
\be
u_z (x,t=0) = U \sin (k x) ,
\ee
where $u_z$ is the $z$-component of the velocity, $k = 2\pi /L$ , $L =
64$ is the linear size of the system, $U=0.07$ is the maximum flow
velocity, $x$ is a Cartesian coordinate, and $t$ is the time.  According
to the Stokes equation, the velocity should evolve as
\be
u_z (x,t) = U \sin ( k x) \exp (-k^2 \nu t)  \; .
\ee
The viscosity can thus be deduced from the integral $\int_0^L \; dx
|u_z| $, which decays exponentially with time.

The minimum viscosity is obtained at a density $d=0.5$ and has the value
$\nu = 0.095$.  For comparison, the value obtained from a purely random
table is $\nu = 0.1667$.  When Q is minimized by choosing the collisions
by an exhaustive search in the (severely) restricted space of 24-bit
{\em isometries}~\cite{henon87b}, the resulting viscosity is within 20\%
of the value resulting from the random table \footnote{In this case the
problem is slightly different because the minimization is of a Reynolds
coefficient rather than of the viscosity.}.

\section{Test simulations}
To check the behavior of the model when the fluid is forced in the
presence of solid walls we performed simulations of flow through a pipe
and channel as well as flow through a simple cubic array of solid
spheres.  The latter geometry creates the only truly three-dimensional flow
because the flow passes through constrictions.
The simulations of pipe flow were carried
out to obtain an estimate of the lower limit for the spatial size of
obstructions and constrictions.
\begin{table}
\begin{center}
\begin{tabular}{|| l |l| l | l |l ||} \hline
Geometry & $\phi $ &  $R$ & $(\kappa -
\kappa_{theory})/\kappa_{theory}$
& $R_{eff}$\\
\hline
\hline
Channel & 1.0 &   1.5     &    +0.04 $\pm $ 0.03  & 1.6   \\
        & 1.0 &   2.5     &    $-$0.04 $\pm $ 0.02  &  2.4\\
        & 1.0 &   3.5     &    $-$0.03 $\pm $ 0.05  & 3.4 \\
        & 1.0 &   7.5     &    +0.03 $\pm $ 0.05    &  7.7\\
        & 1.0 &  15.5     &    +0.04 $\pm $ 0.05  *  & 16.1\\
\hline
Pipe    & 1.0 &   1.5     &    +0.09 $\pm $ 0.07  & 1.6  \\
        & 1.0 &   2.5     &    +0.01 $\pm $ 0.04  &  2.5 \\
        & 1.0 &   8.5     &    $-$0.03 $\pm $ 0.02 &  8.4 \\
        & 1.0 &  15.5     &    $-$0.07 $\pm $ 0.05 *& 14.9 \\
\hline
Array of spheres
        & 0.713 & 4.5     &    $-$0.07 $\pm $ 0.08 &  4.6 \\
        & 0.907 & 4.5     &    $-$0.11 $\pm $ 0.02  & 4.7 \\
        & 0.719 & 6.5     &    $+$0.05 $\pm $ 0.06  & 6.4 \\
        & 0.882 & 19.5    &    $-$0.12 $\pm $ 0.04 * & 20.4\\
        & 0.882 & 19.5    &    $-$0.04 $\pm $ 0.02 & 19.8\\
        & 0.735 & 25.5    &    $-$0.04 $\pm $ 0.07 * & 25.8\\
\hline
\end{tabular}
\end{center}
\caption{
Comparison of calculations of the permeability $\kappa$ to theoretical
predictions $\kappa_{theory}$ for channels, pipes, and a periodic cubic
array of spheres.
$R$ denotes the radius of the pipes and spheres and
the half-width of the channels, $R_{eff}$ denotes
the effective radius (half-width) corresponding to the theoretical
expressions for $\kappa_{theory}$; $\phi$ is the void fraction.
The asterisk (*) shows the simulations that were performed on CAM-8.
(The remaining small-scale simulations were carried out on a workstation.)
}
\end{table}

Table 1 summarizes the results of the simulations of flow through
channels of half-width $R$, pipes with radius $R$, and periodic cubic arrays
of spheres with radius $R$.
The pipes were constructed with periodic boundary conditions in the flow
direction, whereas the channels had periodic boundaries in two directions
and a flat-wall boundary perpendicular to the third direction.
In the table, the permeability $\kappa$ of each particular geometry
is compared to the theoretical prediction of the permeability,
$\kappa_{theory}$.
The permeability is defined as
\be
\kappa = \frac{J \nu \phi}{F}
\ee
where $J$ is the total massflux, $F$ is the total force applied to the fluid,
and $\phi $ the fraction of void space, i.e., the porosity.  Note that
$\phi = 1$ for the pipe and the channel.
In the simulations the permeabilities were deduced from the measurement
of the average flow rate and the average forcing of the fluid, and the
uncertainties are estimated from the noise in these data.

The theoretical expression for the permeability of the pipe,
\begin{equation}
\kappa_{theory} = \frac{R^2}{ 8} \; ,
\end{equation}
is easily obtained from the Stokes equation, as is
the permeability of a channel
\begin{equation}
\kappa_{theory} = \frac{R^2}{3} \; .
\end{equation}
The theoretical expression for the permeability of the simple cubic
array of spheres is obtained (in a slightly different form)
by Hasimoto~\cite{hasimoto59} and Sangani et al.~\cite{sangani82}, and
has the form
\begin{equation}
 \kappa_{theory} = \frac{2}{9 K(\phi )}
 \left( \frac{\phi}{1-\phi} \right) R^2
\end{equation}
where $K(\phi )$ is given in~\cite{sangani82}.  In this case the radii
$R=19.5$ and $26.5$ correspond to the porosities $\phi = 0.882$ and
$\phi = 0.735$ respectively.
The drag $F$ on a single sphere in the
array is
\begin{equation}
F = K(\phi )  \; 6 \pi \rho \nu R \; ,
\label{drag}
\end{equation}
where $\rho $ is the mass density and $\nu$ is the viscosity of the
fluid.  It is seen that $K(\phi ) \geq 1$ is the correction factor to
Stokes law. When the spheres are infinitely far apart $\phi =1 $ and $K=
1$.

The non-integer values of $R$ are discussed below.
They are related to the
definitions of the (discretized) geometries in the following way:
For the pipe a site will be a void site if its distance
$r$ from the symmetry axis satisfies the inequality
\begin{equation}
r < R+0.5 \; .
\end{equation}
The remaining sites on the lattice will be solid sites.
The solid sites of the sphere are given by the same inequality
if $r$ is taken to be the distance from the center of the sphere.

The effective radius $R_{eff}$ given in the above table is the radius
(half-width) that gives the observed permeability when inserted in the
theoretical expressions for  $\kappa_{theory}$. Especially in the case
of the spheres the permeability varies strongly with
$R$, and the difference between $R$ and $R_{eff}$ corresponding
to the discrepancies between $\kappa$ and $\kappa_{theory}$
is small.

The discrepancies between the theoretical and simulated
results have three principal causes. These are
\begin{itemize}
\item Uncertainty in the effective position of the wall;
\item Discretization errors; and
\item Effects of compressibility.
\end{itemize}
In the case of the channel and pipe simulations, only the first and
second problems come into play, whereas all three cause errors in the
three-dimensional case of the spheres.  The discretization errors are
due both to the approximations made when smoothly curved surfaces are
represented on a lattice, and to non-hydrodynamic effects at sufficiently
small scales.

For the hydrodynamic behavior of the lattice gas to be described
by the equations of incompressible hydrodynamics, several requirements
must be fullfilled.
First, the geometry must be such that
there is  a scale separation between
the lattice constant and the scale over which the flow varies.
The ratio between these scales is denoted by the small number
$\epsilon$.
Second,  the flow must be forced sufficiently weakly so that
the flow velocity $u$ does not exceed $\epsilon$.
Finally,  density
variations must  remain small. More precisely, the variations in $\rho$
must be of order $\epsilon^2$ or smaller.
In geometries where the overall permeability is small, a relatively
large force will be necessary to drive the flow
and spatial variations in the permeability (caused, for instance, by
constrictions) may result in large
density variations, even when $u$ is small.

This is the case in particular for the CAM8-simulation with the
$R=19.5$ sphere
where the forcing applied produced a maximum flow velocity $u=0.10$.
This simulation was repeated on a workstation with approximately half the
forcing, and the resulting discrepancy between theory and simulations
was correspondingly decreased.

When the requirements on the scale separation
are obeyed and there is no
forcing, the lattice
gas is described by the following equation~\cite{doolen91}
\begin{equation}
  \frac{\partial {\bf u}}{\partial t} +
  g(\rho ){\bf u}\cdot\nabla {\bf u} =
  -\frac{\nabla P}{\rho} +
  \nu \nabla^2 {\bf u}
  \label{NS}
\end{equation}
where ${\bf u}$ and $\rho$ are the average flow velocity and the
density, respectively,
\begin{equation}
P = (1/2) \rho \; (1- g(\rho )  u^2 )
\label{P}
\end{equation} is the pressure ~\cite{doolen91}
and $\nu$ is the kinematic viscosity.
The $g$-factor is given as
\begin{equation}
g(\rho ) = \frac{4}{3} \left(\frac{12- \rho}{24 - \rho}\right)
\label{g} \; .
\end{equation}
The velocity dependence in the pressure results from the discreteness of
the velocities and the exlusion principle.  For the purpose of finite
Reynolds number simulations, the $g$-factor may be removed by rescaling
the velocity as $u' = g(\rho ) \; u$ \cite{doolen91}.  In the
simulations the average density $\overline{\rho } = 12$ was chosen so
that $g = 0$ and the flow is described by the Stokes equation in the
limit where there are no density variations.

As an example of how to quantify the density variations in steady state
flow, we give an estimate of the density variations that will result in
the case of the array of spheres.  By the above equation of state
(\ref{P}) we have that the density difference across the sphere is given
as
\begin{equation}
\Delta \rho \approx   2 \Delta P \; .
\end{equation}
But this  pressure difference
can be estimated as the
drag divided by some area which is characteristic
for the sphere, say $\pi R^2$.
Hence, by Eq.~(\ref{drag})
\begin{equation}
\Delta \rho \approx   K(\phi) \; 12 \nu \; U/R \; .
\end{equation}
Since the drag coefficient $K( \phi )$ increases sharply
with decreasing porosity $\phi $, $\Delta \rho $ may become
significant even when $U/R \sim \epsilon^2 $ is small.
Care must be taken in order for this  effect of compressibility
in the lattice gas to be negligible.  In the present case when $\rho$
is chosen so that $g(\rho ) =0$, and the Stokes equations
are simulated, deviations in $\rho$ from its average value
will cause a finite value of $g(\rho )$.
This means that the $g$-dependent term of Eq.~(\ref{NS})
must be taken into account and we are no longer in the regime of
linear hydrodynamics.

The effective wall positions have been studied theoretically by several
authors~\cite{cornubert91,ginzbourg94}.  It can be shown (theoretically)
that in the case of Couette flow along a flat wall oriented in a
direction parallel to the lattice directions, the position of vanishing
velocity is not on the wall sites, but rather half a lattice unit
measured from the wall sites into the fluid region.  In the case when
the velocity field has a non-vanishing second derivative, there is a
correction to the effective wall position of relative order
$\epsilon$. This correction is
significant only when the constrictions in the flow are very small, and
it depends on the viscosity.

In the channel flow simulations referred above, we fitted the velocity
profiles with a parabola, as predicted by the Stokes equation. Within
the noise of the measurements we found agreement between simulations and
a parabolic profile going through the position halfway between the first
wall site and the first fluid site.  We also found good agreement when
the wall-normal was in an angle of $\pi /4$ to the closest lattice
directions.  This result slightly generalizes the predicted result for
the wall positions~\cite{ginzbourg94}.  In the table the values of
$R$ are obtained by assuming that the effective wall position is allways
halfway between the first wall site and the first fluid site. In the
case of curved geometries this approximation creates discrepancies
in addition to the discretization approximation.

The relatively good agreement between theory and simulations in the
case of the two smallest pipes is coincidential, since the
crossection of these pipes are squares rather than circles.
In simulations of the flow around the periodic
array of spheres (in the simulations, a single sphere and periodic
boundary conditions were used) the result is highly sensitive to the
exact position of the boundary, as can be seen from the
values of $R_{eff}$ given in the table, and the small discrepancy between
simulation and theory can be accounted for by a correspondingly small
shift of the effective boundary.

In general, the lower limit on the size of obstructions
and constrictions depends on
the particular application as well as the required precision.  In the
case of flow in porous media, the coarse grained characteristics depend
mainly on the flow in the widest channels, and the averaged behavior
depends only weakly on the flow in the narrower passages through the
medium. In this case the small scale hydrodynamics may not matter much.
But the permeability typically depends strongly on the effective
positions of the walls, and care must be taken to ensure that these are
correctly represented or that the channels are sufficiently wide.

\section{Conclusion}
We have designed an algorithm for the implementation of a
three-dimensional (FCHC) lattice gas model on the CAM-8.  It has been
shown how a proper geometric grouping of the velocity vectors on the
lattice makes it possible to decompose both the collisions and the
random isometries acting on the full 24-bit states into operations
involving only 16 or 8 bits at a time. The corresponding 16-bit
collision table has been optimized to obtain a minimum viscosity, and
tested against analytic results from the Boltzmann theory.  The
architecture of the CAM-8 has been described, and the performance of an
eight node machine has been shown to be comparable to that of existing
supercomputers.  Some practical limitations of the model have been
established and discussed, and it has been shown that the model behaves
according to the hydrodynamic predictions for various permeable media.
We have thus demonstrated that this implementation of the FCHC lattice
gas represents a working tool for large-scale simulation of flows in
simple and complex geometries.

\section{Acknowledgements}
We thank John Olson for helpful discussions.  One of us (BMB) would like
to acknowledge helpful conversations with David Levermore and Washington
Taylor.  This work was supported in part by the sponsors of the
MIT Porous Flow Project
and NSF Grant 9218819-EAR.  E. G.  Flekk\o y acknowledges support by
NFR, the Norwegian Research Council for Science and the Humanities.
The CAM-8 project is supported by ARPA under grant N00014-94-1-0662.

\newpage
\clearpage

\clearpage

\clearpage

\clearpage

\end{document}